\documentclass[a4paper]{cas-dc}

\usepackage[numbers,sort&compress]{natbib}
\usepackage{graphicx}
\usepackage{subfigure}
\usepackage{diagbox}
\usepackage{amsmath}
\usepackage{multirow}
\usepackage{amsfonts}

\begin{document}
\let\WriteBookmarks\relax
\def\floatpagepagefraction{1}
\def\textpagefraction{.001}
\shorttitle{Preprint submitted to}
\shortauthors{Wen et~al.}
\shorttitle{}

\title [mode = title]{A multi-level approach with visual information \\for encrypted H.265/HEVC videos}
\author[1]{Wenying Wen}[type=editor,
                        auid=000,bioid=1
                        ]



\author[1]{Rongxin Tu}[style=chinese]
\author[2]{Yushu Zhang}[style=chinese]
\cormark[1]
\ead{e-mail: yushu@nuaa.edu.cn}
\author[1]{Yuming Fang}[style=chinese]
\author[1]{Yong Yang}[style=chinese]
\address[1]{School of Information Management, Jiangxi University of Finance and Economics, Jiangxi, 330013, China.}
\address[2]{College of Computer Science and Technology, Nanjing University of Aeronautics and Astronautics, Nanjing 211106, China}








\begin{abstract}
High-efficiency video coding (HEVC) encryption has been proposed to encrypt syntax elements for the purpose of video encryption. To achieve high video security, to the best of our knowledge, almost all of the existing HEVC encryption algorithms mainly encrypt the whole video, such that the user without permissions cannot obtain any viewable information. However, these encryption algorithms cannot meet the needs of customers who need part of the information but not the full information in the video. In many cases, such as professional paid videos or video meetings, users would like to observe some visible information in the encrypted video of the original video to satisfy their requirements in daily life. Aiming at this demand, this paper proposes a multi-level encryption scheme that is composed of lightweight encryption, medium encryption and heavyweight encryption, where each encryption level can obtain a different amount of visual information. First, we use AES-CTR to generate a pseudo-random number sequence. Then, the main syntax elements in the H.265/HEVC encoding process are encrypted by a pseudo-random sequence. In the lightweight encryption level, the syntax element of the luma intraprediction model is chosen for encryption. In the medium encryption level, the syntax element of the DCT coefficient sign is employed for scrambling encryption. In the heavyweight encryption level, syntax elements of both the luma intraprediction model and the DCT coefficient sign are encrypted simultaneously by the pseudo-random sequence. It is found that both encrypting the luma intraprediction model (IPM) and scrambling the syntax element of the DCT coefficient sign can achieve the performance of a distorted video in which there is still residual visual information, while encrypting both of them can implement the intensity of encryption and one cannot gain any visual information. The experimental results meet our expectations appropriately, indicating that there is a different amount of visual information in each encryption level. Meanwhile, users can flexibly choose the encryption level according to their various requirements.

\end{abstract}

\maketitle

\section{Introduction}

High-efficiency video coding (HEVC) \cite{grois2013performance} is the latest video coding standard that was published by ISO/IEC MPEG, and ITU-T VCEG formed the Joint Collaborative Team on Video Coding (JCT-VC) in 2013, which has a high efficiency to compress video. HEVC is adapted to the transmission and storage from small-scale multimedia networks to large scale TV distributors and thus has been widely used in daily life \cite{Li2020FastDI,Sasithradevi2020VideoCA,8924951}. Video contains an enormous amount of information including private, sensitive and copyright items \cite{wen2020colour,Sridhar2018AWB}, which would be easily leaked in an unreliable public channel and the insecurity of the cloud service. Currently, video encryption has been a challenging research topic, as a technology applied in military, medical and other related industries to maintain data security.

Video encryption provides a secure channel during transmission. To the best of our knowledge, in most of the existing video encryption schemes, the whole video is encrypted, such that the user without permissions cannot obtain any viewable information. A user with the secret key can see the original video, whereas users without a key cannot receive any visual information. However, there are not enough choices to meet the needs of people with a variety of demands. For example, in the professional paid video scenario, users have to pay an expensive fee to watch the video; otherwise, they cannot see any video information. A large number of people need the professional video, but they cannot afford the expensive cost. If the encryption video can be divided into a multi-level approach where the different levels have different amounts of video information, then the video provider can set multiple charging standards according to the amounts of video information.
 To some extent, this alleviates the problem of supply and demand in the market, so it is good for both the user and provider. Another instance occurs in important video meetings, such that if we can divide the encryption meeting video into a multi-level approach with different levels that contain different amounts of information, the leader can set multiple grades of the users according to the amounts of visual information in the video. Only in this way will the meeting video be even more effective in people's work and lives \cite{Duvar2019FastIM,Peng2019MultipleCF}.

 \begin{table*}
 \renewcommand\arraystretch{1.1}
\normalsize
\centering
\setlength{\tabcolsep}{4mm}
\caption{Description of symbols used in the paper}
\begin{tabular}{lll}
\hline\hline
\textit{\textbf{Symbols}} &  & \textit{\textbf{Definition}}                                                             \\ \hline
HEVC, AES            &  & High-Efficiency Video Coding, Advanced Encryption Standard \\
DES, IDEA              &  & Data Encryption Standard, International Data Encryption Algorithm             \\
PM, IPM, MV            &  & Prediction Modes, Intraprediction Modes, Motion Vectors               \\
MI, MVD, TC         &  & Merge Index, Motion Vectors Difference, Transform Coefficients \\
RPS, QP, MVP                 &  & Reference Picture Set, Quantized Transform, Motion Vector Prediction               \\
RefFI, NEA                 &  &  Reference Frames Index, Naïve Encryption Algorithms \\
SE, DC       &  & Selective Encryption Algorithms, Discrete Cosine \\
DCT, CAVLC                 &  & Discrete Cosine Transformation, Context-Adaptive Variable Length Coding                                            \\
PSNR, SSIM              &  & Peak Signal-to-Noise Ratio, Structural Similarity  \\
NPCR, UACI                &  & Number of Pixel Change Rate, Uniform Average Change Intensity                               \\ \hline
\end{tabular}
\label{1}
\end{table*}

This paper proposes a multi-level encryption approach based on AES, and then a tunable selection encryption scheme can meet the various requirements of users. First, we use AES-CTR to generate a pseudo-random number sequence. Then, the main syntax elements in the H.265 /HEVC encoding process are encrypted by a pseudo-random sequence. In the process, only one syntax element is encrypted at a certain encryption level \cite{Yao2015ALA}. It can be seen that the encryption of the luma intraprediction model (IPM) and the scrambling of the DCT coefficient sign can achieve multi-level encryption with different visual information. Different levels of the encryption video can adapt to various application scenarios that depend on the requirements of the users.

The main contributions of this paper include the following points:

\begin{enumerate}
\itemsep=0pt
\item {} A multi-level selection scheme for encrypted H.265/H-EVC is proposed. It divides the encryption video into three levels. In video frames with the lightweight encryption level, the luma IPM in table \ref{1} is merely encrypted. In video frames with the medium encryption level, the DCT coefficient sign is chosen for encryption. Moreover, in the heavyweight encryption level, both the luma IPM and DCT coefficient sign are employed for encryption.

\item {} In the proposed scheme, different levels contain different amounts of visual information.  In the first two levels, the object features and the outline and structure information of the object are identified, while no useful information can be gained from the encrypted video in the last level. The proposed multi-level encryption approach for H.265/HEVC is provided to meet various scenario application requirements, and it exhibits the flexibility of the proposed algorithm.

\item {} We have theoretically analysed and experimentally tes-ted the performance of the encryption of the luma IPM and DCT coefficient sign. It can be found that the encryption of both syntax elements greatly distorts the video, while encrypting these syntax elements individually would reserve different kinds of visual information.
\end{enumerate}

The rest of the paper is organized as follows. In Section 2, the related work of encryption video is introduced. Preliminary knowledge of the HEVC framework and AES algorithm are introduced in Section 3. The proposed multi-level encryption scheme is provided in Section 4. The experimental results and security analysis are depicted in Section 5. Finally, the conclusions are given in Section 6.

\begin{table}
\renewcommand\arraystretch{1.1}
\centering
\setlength{\tabcolsep}{0.1mm}
\caption{The characteristics of the related H.265/HEVC encryption algorithms}
\begin{tabular}{llc}
\hline \hline
\textit{\textbf{\begin{tabular}[c]{@{}l@{}}Encryption \\ scheme\end{tabular}}} & \textit{\textbf{\begin{tabular}[c]{@{}l@{}}Encryption \\ elements\end{tabular}}}                                                          & \multicolumn{1}{l}{\textit{\textbf{\begin{tabular}[c]{@{}l@{}}Entirely \\ encryption\end{tabular}}}} \\ \hline
Qiao{\cite{qiao1997new}}                                                                   & Bit Stream                                                                                                                                & Yes                                                                                                  \\\hline
Jolly{\cite{shah2011video}}                                                                    & Bit Stream                                                                                                                                & Yes                                                                                                  \\\hline
Lian{\cite{1649688}}                                                                     & TC and MVD Sign                                                                                                                           & Yes                                                                                                  \\\hline
Wallendael{\cite{van2013format}}                                                                                 &IPM, RFI &  Yes  \\\hline
Wallendael{\cite{van2013encryption}}                                                               & RPS, QP, Residual Sign and SAO                                                                                                            & Yes                                                                                                  \\\hline
Peng{\cite{peng2015effective}}                                                                   & \begin{tabular}[c]{@{}l@{}}Residual Sign, RFI, DC Coeff Sign,\\ MVD Sign and Value\end{tabular}                                            & No                                                                                                   \\\hline
Wang{\cite{wang2010energy}}                                                                     & IPM, Inter-PM                                                                                                                             & No                                                                                                   \\
\hline
Zhao{\cite{6335618}}                                                                     & Compressed Domain                                                                                                                         & Yes                                                                                                  \\
\hline
V.A.Memos{\cite{Memos2015Encryption}}                                                                & Residual Coefficients of I Frame                                                                                                          & Yes                                                                                                  \\
\hline
Boyadjis{\cite{boyadjis2016extended}}                                                                 & \begin{tabular}[c]{@{}l@{}}Luma IPM, SAO, MVPIdx,\\ MVD Sign and Value\end{tabular}                                                        & Yes                                                                                                  \\\hline
Peng{\cite{peng2019tunable}}                                                                   & \begin{tabular}[c]{@{}l@{}}Luma IPM, Chroma IPM, RefFI, MI,\\  MVD Sign and Value, MVP Index, \\ Residual Sign and Value, SAO\end{tabular} & Yes                                                                                                  \\ \hline
\end{tabular}
\label{2}
\end{table}

\section{Related Work}

In past encryption schemes, a code video is regarded as a bit stream, and some traditional ciphers such as the Advanced Encryption Standard (AES) \cite{Sklavos2002Architectures} or other bit stream ciphers are used to encrypt the bit stream. The method is called naive encryption algorithms (NEA). The idea of the NEA does not apply to any of the syntax elements and special structures but just treats the HEVC stream as text data. There is no existing algorithm that can break triple AES, so it can provide high security to the video because each byte is encrypted. In \cite{qiao1997new}, NEA MPEG videos were proposed by Qiao \begin{itshape}et al\end{itshape}. MPEG encoded video is encrypted in each byte by the International Data Encryption Algorithm (IDEA), which is used to generate a pseudo-random sequence. In \cite{shah2011video}, J. Shah \begin{itshape}et al\end{itshape}. proposed a scheme by using DES and AES to encrypt the bit steam of MPEG video. Both of their schemes can provide a high security to protect the video that is guaranteed by AES \cite{Sklavos2002Architectures} or DES \cite{Ehrsam1978A}. However, they are not applications suitable for large videos because the speed of encryption is very slow.
Moreover, using the NEA to encrypt the video with high resolutions can result in high computational complexity, which makes it impossible to meet the requirements of real-time transmission \cite{Hamidouche2017RealtimeSV}. Therefore, selective encryption (SE) algorithms \cite{tew2015hevc,Sallam2018HEVCSE,Yang2017AnEL,Liu2019TraceablethenrevocableCA,Maniccam2001LosslessIC} for video have been attracting much attention.

\begin{figure*}
\centering
\includegraphics [scale=0.26]{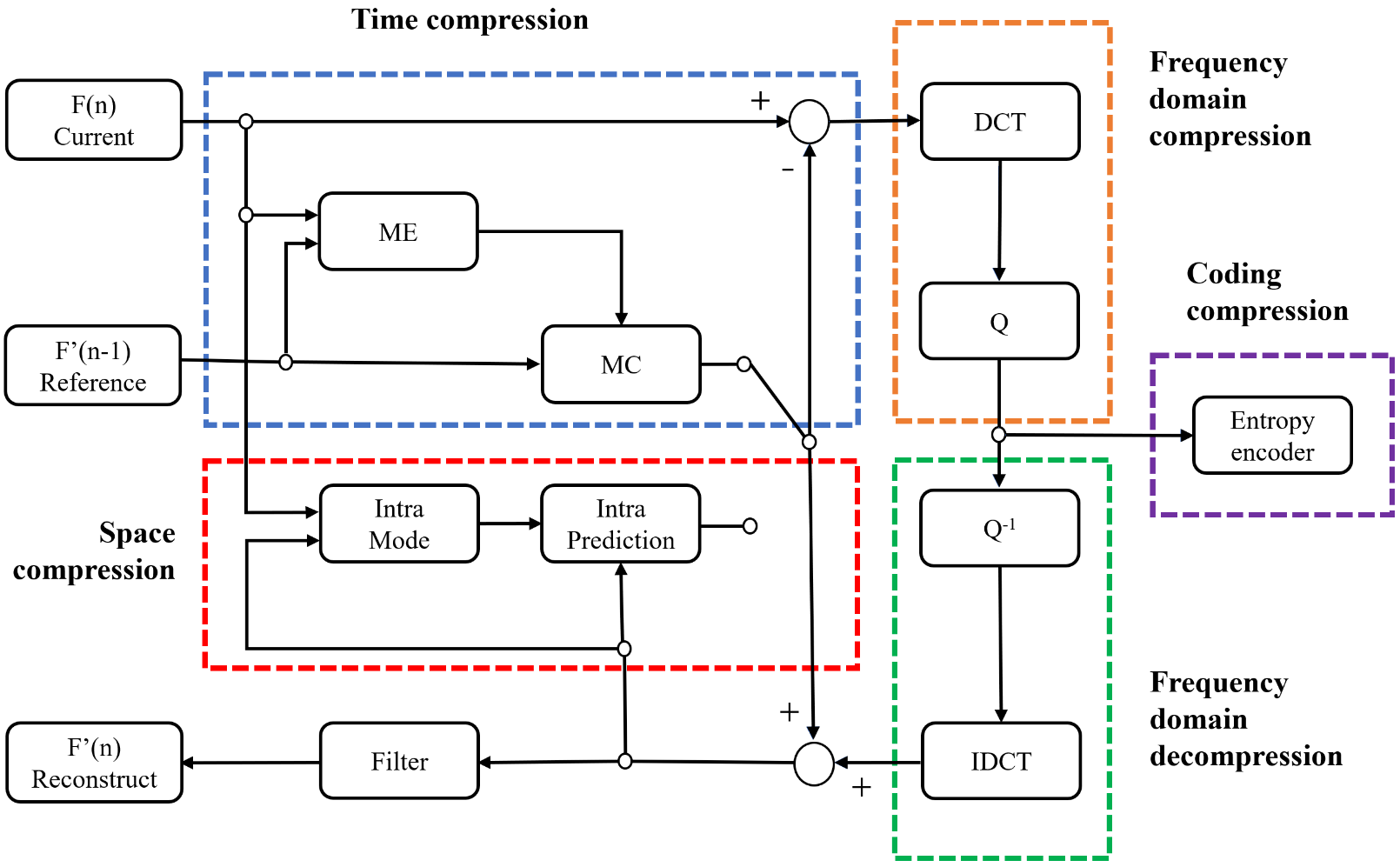}
\caption{The framework of H.265/HEVC.}
\label{I}
\end{figure*}

In the video coding process, some syntax elements play a very significant role that can affect the quality of the final encoding video. Selective encryption algorithms usually encrypt some important syntax elements in the coding process, and the standard decoder can decode the encrypted video. Nevertheless, after decoding, encrypted video is seriously distorted so that one cannot obtain any useful information, and users with a secret key can acquire the original video. The SE scheme for the H265/HEVC stream was exploited from the work of Lian \begin{itshape}et al\end{itshape}. \cite{1649688}, which proposed the encryption of the syntax elements, intraprediction modes from transform coefficients and motion information to distort the video. In 2013, the new standard of the H265/HEVC was published. Wallendael \begin{itshape}et al\end{itshape}. \cite{van2013format} proposed an extensive encryption scheme by selecting some syntax elements in the encoding process for encryption, which included Intra syntax elements and Inter syntax elements in the H265/HEVC stream. Simultaneously encrypting these syntax elements can distort the video frame to achieve the effect of encryption. In \cite{van2013encryption}, Wallendael \begin{itshape}et al\end{itshape}. involved more syntax elements for encryption including reference picture set (RPS), quantization parameter (QP), inter-frame information, residual information, de-blocking and sample adaptive offset parameters. The experimental results indicate that the encryption of these syntax elements further can enhance the encryption effect of the video. An SE algorithm was proposed by Peng \begin{itshape}et al\end{itshape}. \cite{peng2015effective}, who used the Rossler chaotic system to generate a pseudo-random sequence to encrypt many syntax elements. Even though the scheme has a good encryption performance, the bit rate of the coding generally increases. Wang \begin{itshape}et al\end{itshape}. \cite{wang2010energy} proposed a method that considered the relationship between the current and descendant frames that encrypted current frames more dependent on descendant frames. They just encrypted the current frames, while the dependent frames are not encrypted. Therefore, this method reduces the bit rate of the video to a large extent. Zhao \begin{itshape}et al\end{itshape}. \cite{6335618} divided the video frames into foreground and background and then only encrypted the foreground that contains important information. Peng \begin{itshape}et al\end{itshape}. \cite{peng2013roi} employed a protection scheme based on FMO and chaos for the regions of interest (ROI), which provided a low bit rate of the video. V. A. Memos \begin{itshape}et al\end{itshape}. \cite{Memos2015Encryption} proposed an unequal scheme that selected the residual coefficients of the I frame to encrypt. It also has a good performance in visual distortion because B frames and P frames are predicted by the I frame. However, the encryption space is small and the security encryption is insufficient.
 Boyadjis \begin{itshape}et al\end{itshape}. \cite{boyadjis2016extended} presented a method to encrypt the syntax elements such as the luma intraprediction mode and quantized transform coefficients. The information of edge regions is not given much consideration, though the encryption performance of the I frame is enhanced. Peng \begin{itshape}et al\end{itshape}. \cite{peng2019tunable} extended this technique such that they encrypted the edge regions by scrambling coefficients based on edge extraction. To enhance encryption performance, they further encrypted the chroma intraprediction mode. The distortion of video achieved great improvement.

\begin{figure*}
	\centering
    \includegraphics [scale=0.3]{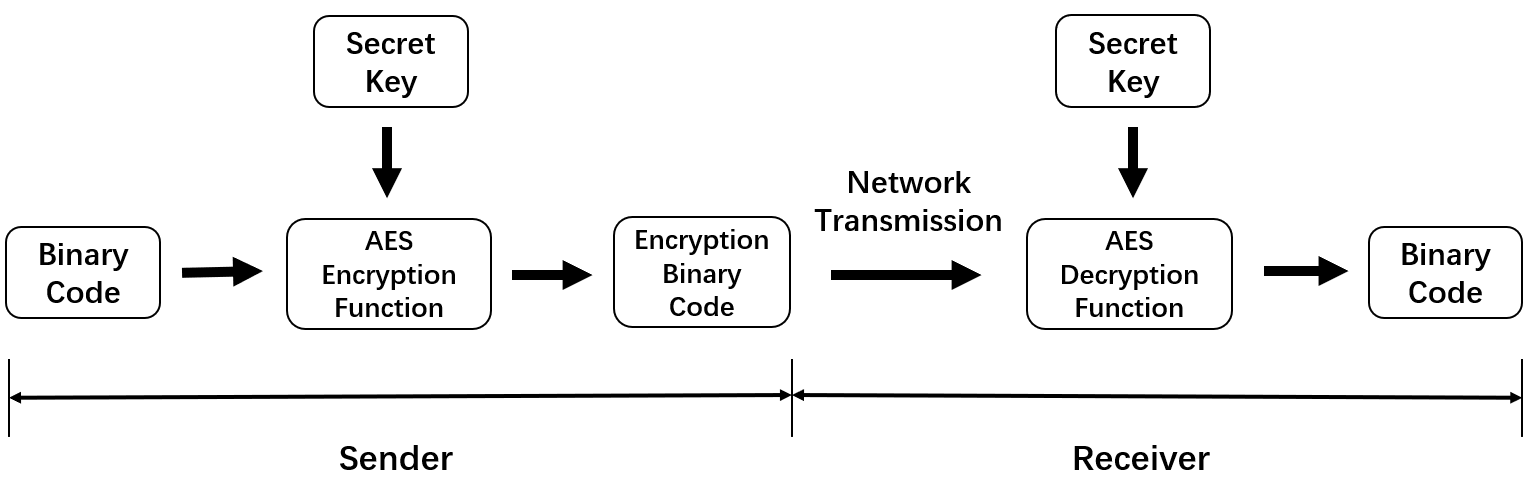}
	\caption{The diagram of the AES encryption algorithm.}
	\label{II}
\end{figure*}
\vspace{-0.1cm}
However, most of the aforementioned video encryption algorithms focus mainly on the whole video encryption that depends on the syntax elements and the whole video to be encrypted. The authorized persons can obtain the original video, while an unauthorized user gains an encryption video without any useful information. There are not enough choices provided for users when the whole video is encrypted. Therefore, there are not enough choices to meet the needs of people with varieties of demands. Moreover, the characteristics of the aforementioned H.265/HEVC encryption algorithms are listed in Table \ref{2}. To solve the above-mentioned problems in the exiting video encryption algorithms, a multi-level video encryption scheme based on the AES cipher is proposed in this paper to provide sufficient choices for users.

\section{Preliminary Knowledge}
\begin{enumerate}[3.1]
\item Framework of HEVC

From the perspective of the coding framework, H265/H-EVC and H264/AVC \cite{minemura2016novel,peng2017selective} are overall basically the same. The framework of HEVC is shown in Figure \ref{I}. H265/HEVC has been further optimized in the prediction, transform, quantization, entropy coding \cite{Gao2016AnEE} and filtering processes \cite{shen2016combined}. Thus, HEVC has similar sharpness but needs half of the video bitstream. Essentially, video encoding is a hybrid compression coding algorithm, including intra-frame static compression (the blue dotted box in Figure \ref{I}) \cite{Jaballah2018LowCI}, inter-frame dynamic compression (the red dotted box) \cite{Fu2020EarlyTF}, frequency domain compression (the orange dotted box) and bitstream compression (the purple dotted box), which makes it possible for video to perform storage and transmission \cite{Kim2019TransformWR,Fang2019VideoSD}. There are many syntax elements involved in the compression process and eventually rendered as a bitstream. Moreover, in one frame of the encoding process, due to referring to the inter-frame, the encoder needs a reconstruction frame, which has a reconstruction path and stores the frame in a buffer. The process of decoding is just an inverse of the encoding \cite{de2016ghevc}.

\begin{figure*}
\centering
\includegraphics [scale=0.360]{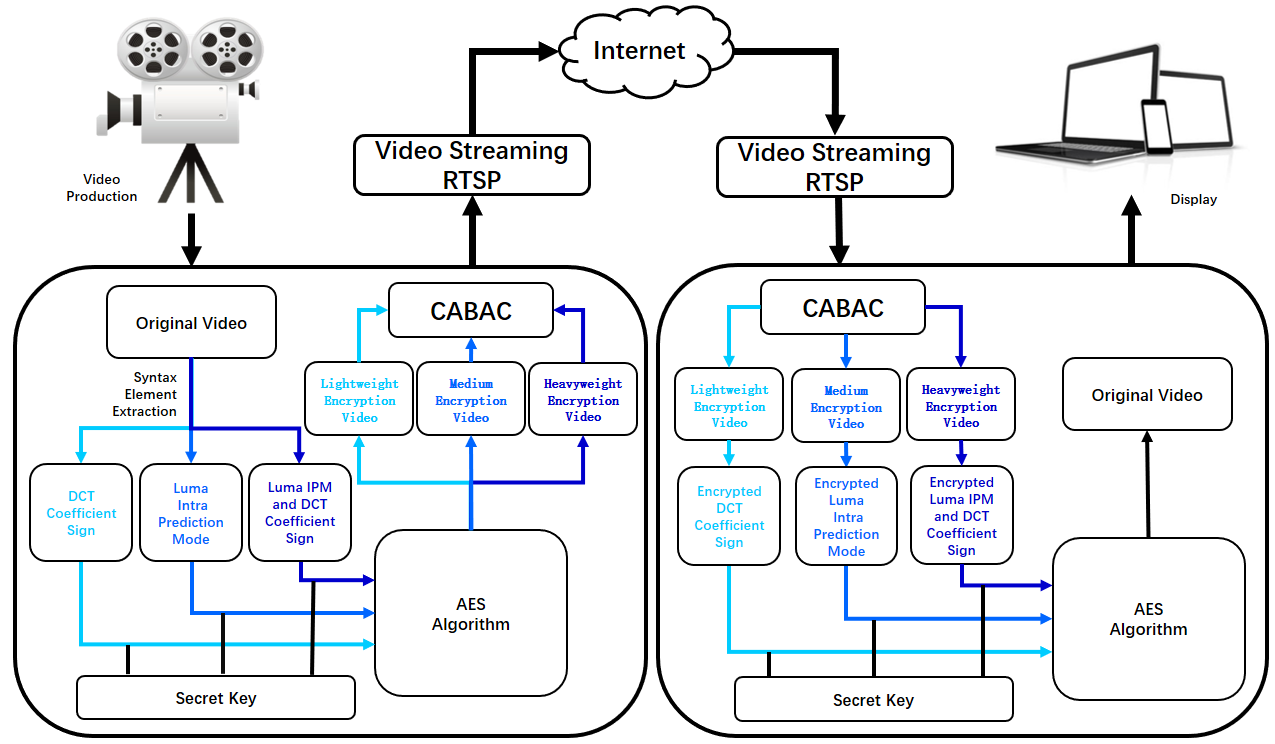}
\caption{The proposed multi-level encryption scheme.}
\label{III}
\end{figure*}

\item AES Encryption Algorithm
\vspace{0.1cm}

AES has high security and is the most common symmetric encryption algorithm, in which the encryption key is the same as the decryption key. The specific encryption process is shown in Figure \ref{II}.
Regarding the AES encryption function as $E$, the encryption process is depicted as follows:

\begin{equation}
C = E (K, P).
\end{equation}
where $K$ is the secret key, $P$ is the binary code, and $C$ is the encrypted binary code. Actually, placing the secret key and binary code into the function $E$, it would output the encrypted binary code.
Regarding the AES decryption function as $D$, the encryption process can be represented as follows:

\begin{equation}
P = D (K, C).
\end{equation}
The decryption process is an inverse of the encryption process.
In this paper, $P$ is the video bitstream by an entropy coding, which is the binary code. Moreover, $C$ is the scrambled entropy coding, which is encrypted in the video.

\end{enumerate}

\section{The Proposed Scheme}
This paper proposes a multi-level encryption method for H265/HEVC. First, we use AES-CTR to generate a pseudo-random number sequence. Then, the main syntax elements in the H.265/HEVC encoding process are encrypted by a pseudo-random sequence. It divides the encryption video into three levels. In video frames with the lightweight encryption level, the luma IPM in table \ref{1} is merely encrypted, and the features of object can identified. In video frames with the medium encryption level, the DCT coefficient sign is chosen for encryption, and the outline and structure information of the object can be identified. Additionally, in the heavyweight encryption level, both the luma IPM and DCT coefficient sign are selected for encryption, and one cannot gain any useful information from the encrypted video. The proposed multi-level encryption approach for H.265/HEVC is provided to meet various scenario application requirements, and it exhibits the flexibility of the proposed algorithm. The framework of the proposed scheme is shown in Figure \ref{III}.

In each encryption level, the related syntax elements have to be encrypted by the AES algorithm. First, we employ AES-CTR with an initial key $N$ to generate a pseudo-random $K$ (the secret key), and it can be depicted as
 \begin{equation}
 K = AES-CTR (N).
\end{equation}
where AES-CTR (-) is operated in counter mode. Through transforming a block cipher to a stream cipher, it generates a pseudo-random sequence by encrypting successive values of arbitrary length. The random sequence is produced by the counter without repeating for a long time. More details of CTR are described in the work \cite{lipmaa2000ctr}.

Then, we use the generating pseudo-random sequence K to encrypt each binary syntax element. The length of K depends on the encryption syntax elements. The encryption process is represented as follows.

\begin{enumerate}[4.1]
\item Lightweight Encryption Level
\vspace{0.1cm}

In the lightweight encryption level, the syntax element of luma IPM is encrypted. In the coding process, luma IPM plays a significant role; B.Boyadjis et al. \cite{boyadjis2016extended} proposed to encrypt luma IPM. Due to the strong correlation between the current coding unit and adjacent pixels in HEVC, the current coding unit is modelled with the encoded pixels. Moreover, it proposes 35 prediction modes (from 0 to 34), including planar, DC and angles. Then, the encoder employs traversal prediction modes in total to determine the minimum rate of distortion as the optimal prediction mode.

\vspace{0.1cm}

 Moreover, the encoder is not directly recoding the optimal prediction mode that needs a 5-bit offset DIR but first establishes a candidate mode list of 3 bits according to the neighbouring Pus because the current coding unit has a very high probability of being the same as the neighbouring Pus. If the current intraprediction mode is in the list, then the encoder needs 3 bits but not 5 bits to recode the prediction mode, and to a great extent, it reduces the bit rate and the list number is recorded. The $Idx$ of the list number is encrypted, and it is defined as
\begin{equation}
 En \_ Idx = (Idx + K_i) \% 3  \qquad 0 \leq K_i \leq 3.
\end{equation}
where $K_i$ represents a segment in the pseudo-random sequence $K$. If the current luma IPM is not in the list, then we are going to scramble the optimal prediction mode; then, there is a large probability of obtaining a bad prediction mode that would distort the video in the coding process. The encryption is defined as
\begin{equation}
 En \_ Idx = Idx \oplus K_i  \qquad\quad 0 \leq K_i \leq 31.
\end{equation}
where $\oplus$ represents the $XOR$ operation.
The encryption of luma IPM performs $XOR$ operations between the number of the 5-bit offset or the 3-bit candidate mode list with a secret key. That is, the recoded optimal prediction mode has scrambling to other prediction modes that are not suitable to predict the current coding unit and even has a high probability to obtain a terrible prediction. It leads to a distortion of the decoded video. Actually, only encrypting the luma intraprediction mode cannot achieve full encryption, and the outline information of the objects is still visible.

\vspace{0.1cm}
 However, the encryption video with visual information is the exact requirement for certain application scenarios. This paper sets the lightweight encryption level as the first level by encrypting the luma intraprediction mode.

\vspace{0.1cm}
The luma intraprediction mode should be the same between the process of coding and decoding; otherwise, it will cause decoding failure because different modes need different parameters. The encoder has to set up a new array to record the scrambling list number or bit offset to solve the asynchronous problem between the coding and decoding.

\begin{figure*}
\centering
\begin{minipage}{0.19\linewidth}
  \centerline{\includegraphics[width=3cm,height=2.5cm]{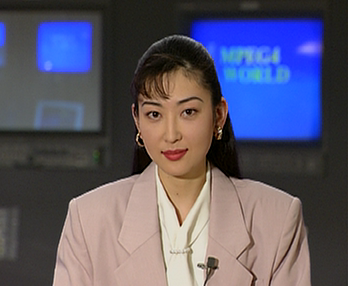}}
  \centerline{(a) }
\end{minipage}
\hfill
\begin{minipage}{.19\linewidth}
  \centerline{\includegraphics[width=3cm,height=2.5cm]{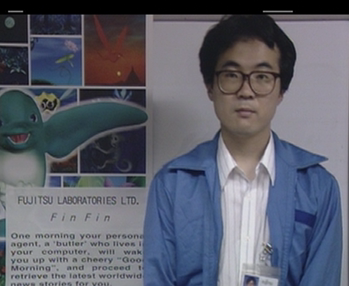}}
  \centerline{(b) }
\end{minipage}
\hfill
\begin{minipage}{.19\linewidth}
  \centerline{\includegraphics[width=3cm,height=2.5cm]{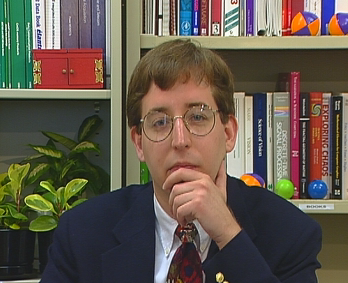}}
  \centerline{(c)}
\end{minipage}
\hfill
\begin{minipage}{.19\linewidth}
  \centerline{\includegraphics[width=3cm,height=2.5cm]{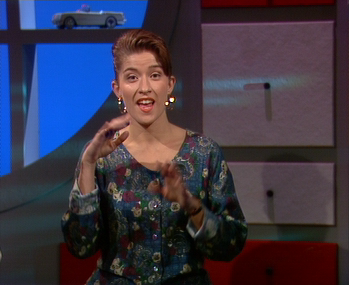}}
  \centerline{(d)}
\end{minipage}
\hfill
\begin{minipage}{.19\linewidth}
  \centerline{\includegraphics[width=3cm,height=2.5cm]{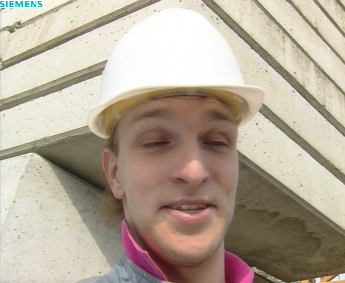}}
  \centerline{(e)}
\end{minipage}

\vfill
\begin{minipage}{0.19\linewidth}
  \centerline{\includegraphics[width=3cm,height=2.5cm]{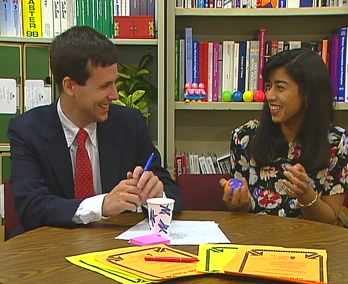}}
  \centerline{(f)}
\end{minipage}
\hfill
\begin{minipage}{.19\linewidth}
  \centerline{\includegraphics[width=3cm,height=2.5cm]{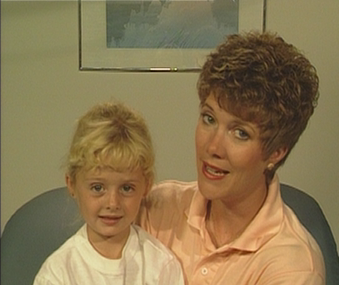}}
  \centerline{(g)}
\end{minipage}
\hfill
\begin{minipage}{.19\linewidth}
  \centerline{\includegraphics[width=3cm,height=2.5cm]{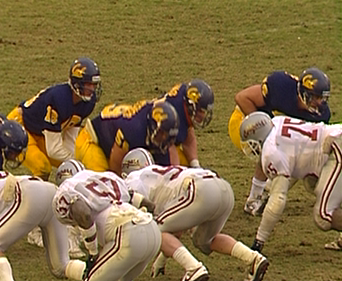}}
  \centerline{(h)}
\end{minipage}
\hfill
\begin{minipage}{0.19\linewidth}
  \centerline{\includegraphics[width=3cm,height=2.5cm]{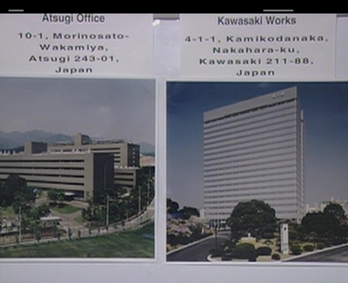}}
  \centerline{(i)}
\end{minipage}
\hfill
\begin{minipage}{0.19\linewidth}
  \centerline{\includegraphics[width=3cm,height=2.5cm]{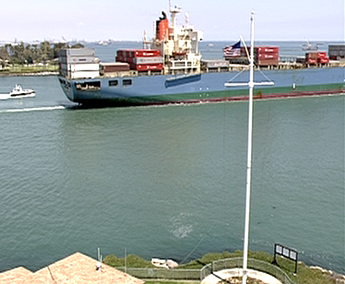}}
  \centerline{(j)}
\end{minipage}
\centering
\caption{Video test sequences. (a) Akiyo. (b) Bowing. (c) Deadline. (d) Irene. (e) Foreman. (f) Paris. (g) mother. (h) Football. (i) pamphlet. (j) Container.}
\label{IV}
\end{figure*}

\vspace{0.1cm}
\item Medium Encryption Level
\vspace{0.1cm}

In the medium encryption level, the syntax element of the DCT coefficient sign is encrypted. In HEVC, for further compression of the video simultaneously without much distortion, it is transformed from the time domain information into the frequency domain information by using DCT. In the frequency domain, the low-frequency signal contains the main information, whereas the high-frequency signal contains the object edge information that generally would be a zero setting due to the small effect on vision. After the discrete cosine transform, the 2-D block of the DCT matrix is converted into a 1-D array by using a scan pattern that defines a processing order for the coefficients. Then, the 1-D array is going to be coding by the context-adaptive variable length coding (CAVLC) \cite{ShahidVisual}. After implementing DCT and quantification, there are many zeros in the array. CAVLC codes the number of zeros, the position, the value and the sign of non-zeros.
 More details of CAVLC are described in the work \cite{ShahidVisual}. In the coding process, TotalCoeffs and TrailingOnes cannot be encrypted because they will lead to decoding failure \cite{Zhou2019SpatialEC}.
In the proposed scheme, the sign of the TrailingOnes is to be encrypted. Coefsign = 1 and coefsign = 0, respectively, represent positive and negative, and the encryption is accomplished to exchange each other. After scanning the DCT matrix, the TrailingOnes values are on the right of the 1-D array that contains the high frequency information. Some details of the figure enhance image sharpness, and then, encrypting the sign of TrailingOnes would not influence the overall outline and acts as a slight perturbation to the image. The encryption of the DCT coefficient sign is represented as
\begin{equation}
 En \_ sign = sign \oplus K_i  \qquad\quad 0 \leq K_i \leq 1.
\end{equation}

Although the syntax element of the video is encrypted, there still exists a large number of visual information. The effect of the encryption video that has visual information is the exact requirement for certain application scenarios. This paper employs the DCT coefficient sign for encryption as the second level, that is, the medium encryption level.

\vspace{0.1cm}
\item  Heavyweight Encryption Level
\vspace{0.1cm}

In the heavyweight encryption level, both the luma IPM and DCT coefficient sign are chosen for encryption. The ways of encryption for the syntax elements are depicted in Section 4.1 and Section 4.2. When the syntax elements are encrypted, one cannot gain any visual information from the video. Both the luma IPM and the DCT coefficient sign are employed to encrypt as the third level, that is, the heavyweight encryption level.

\begin{figure*}
	\centering
	\begin{minipage}{0.19\linewidth}
		\centerline{\includegraphics[width=3cm,height=2.5cm]{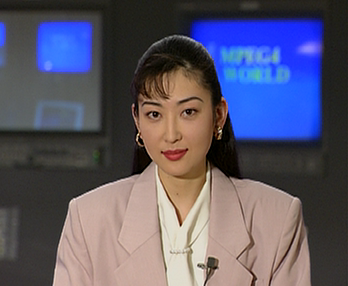}}
		\centerline{(a) }
	\end{minipage}
	\hfill
	\begin{minipage}{.19\linewidth}
		\centerline{\includegraphics[width=3cm,height=2.5cm]{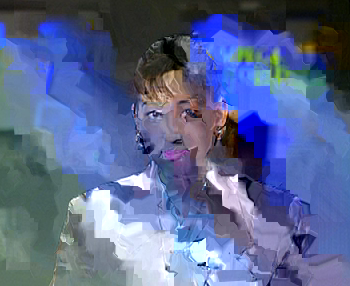}}
		\centerline{(b) }
	\end{minipage}
	\hfill
	\begin{minipage}{.19\linewidth}
		\centerline{\includegraphics[width=3cm,height=2.5cm]{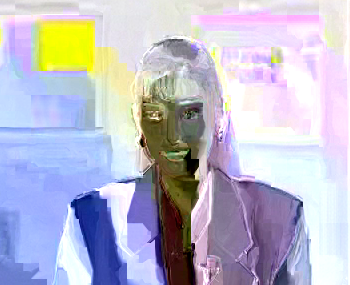}}
		\centerline{(c)}
	\end{minipage}
	\hfill
	\begin{minipage}{.19\linewidth}
		\centerline{\includegraphics[width=3cm,height=2.5cm]{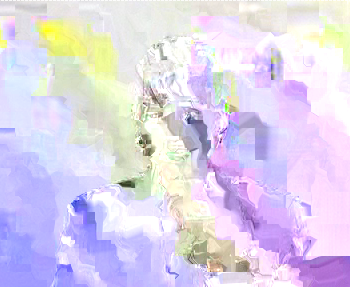}}
		\centerline{(d)}
	\end{minipage}
	\hfill
	\begin{minipage}{.19\linewidth}
		\centerline{\includegraphics[width=3cm,height=2.5cm]{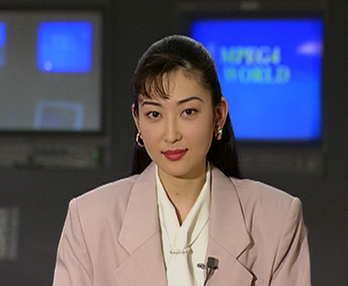}}
		\centerline{(e)}
	\end{minipage}

	\vfill
	\begin{minipage}{0.19\linewidth}
		\centerline{\includegraphics[width=3cm,height=2.5cm]{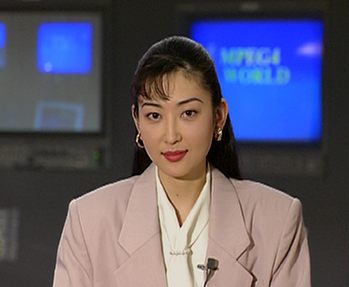}}
		\centerline{(f)}
	\end{minipage}
	\hfill
	\begin{minipage}{.19\linewidth}
		\centerline{\includegraphics[width=3cm,height=2.5cm]{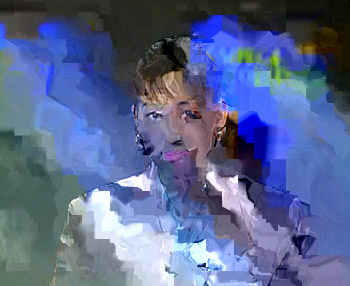}}
		\centerline{(g)}
	\end{minipage}
	\hfill
	\begin{minipage}{.19\linewidth}
		\centerline{\includegraphics[width=3cm,height=2.5cm]{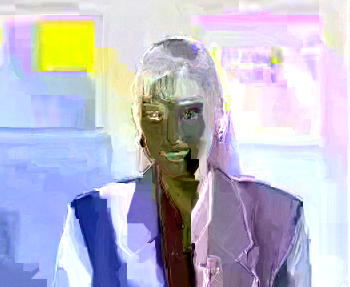}}
		\centerline{(h)}
	\end{minipage}
	\hfill
	\begin{minipage}{0.19\linewidth}
		\centerline{\includegraphics[width=3cm,height=2.5cm]{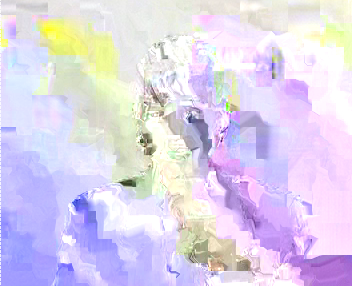}}
		\centerline{(i)}
	\end{minipage}
	\hfill
	\begin{minipage}{0.19\linewidth}
		\centerline{\includegraphics[width=3cm,height=2.5cm]{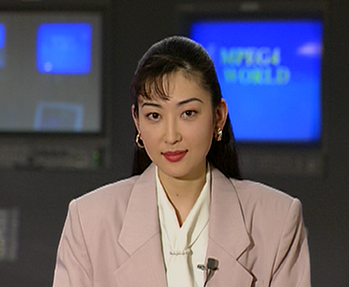}}
		\centerline{(j)}
	\end{minipage}
	
	\vfill
	\begin{minipage}{0.19\linewidth}
		\centerline{\includegraphics[width=3cm,height=2.5cm]{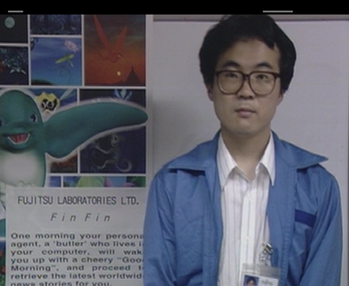}}
		\centerline{(k)}
	\end{minipage}
	\hfill
	\begin{minipage}{.19\linewidth}
		\centerline{\includegraphics[width=3cm,height=2.5cm]{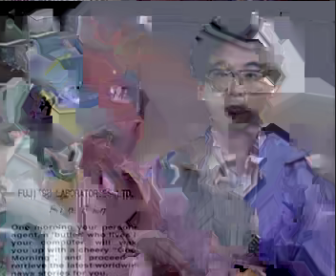}}
		\centerline{(l)}
	\end{minipage}
	\hfill
	\begin{minipage}{.19\linewidth}
		\centerline{\includegraphics[width=3cm,height=2.5cm]{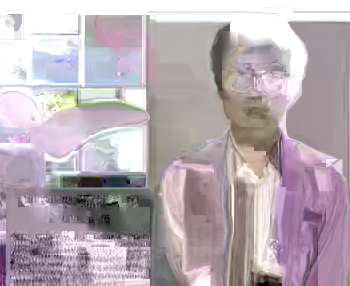}}
		\centerline{(m)}
	\end{minipage}
	\hfill
	\begin{minipage}{0.19\linewidth}
		\centerline{\includegraphics[width=3cm,height=2.5cm]{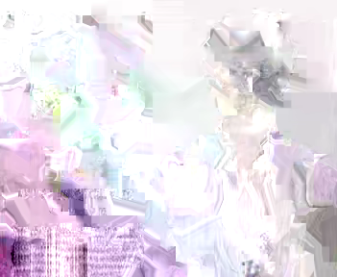}}
		\centerline{(n)}
	\end{minipage}
	\hfill
	\begin{minipage}{0.19\linewidth}
		\centerline{\includegraphics[width=3cm,height=2.5cm]{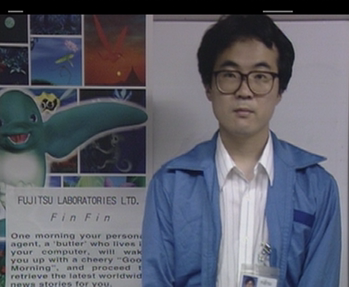}}
		\centerline{(o)}
	\end{minipage}
	
	\vfill
	\begin{minipage}{0.19\linewidth}
		\centerline{\includegraphics[width=3cm,height=2.5cm]{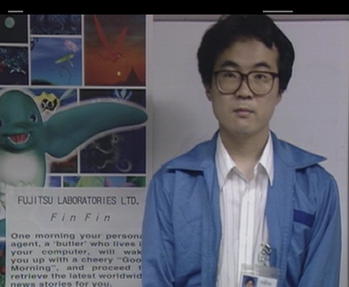}}
		\centerline{(p)}
	\end{minipage}
	\hfill
	\begin{minipage}{.19\linewidth}
		\centerline{\includegraphics[width=3cm,height=2.5cm]{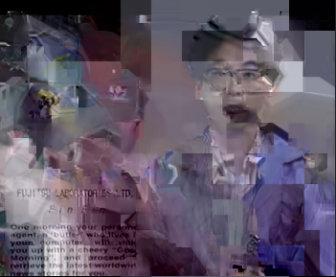}}
		\centerline{(q)}
	\end{minipage}
	\hfill
	\begin{minipage}{.19\linewidth}
		\centerline{\includegraphics[width=3cm,height=2.5cm]{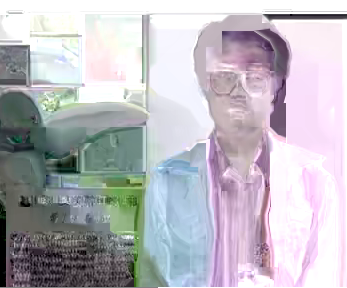}}
		\centerline{(r)}
	\end{minipage}
	\hfill
	\begin{minipage}{0.19\linewidth}
		\centerline{\includegraphics[width=3cm,height=2.5cm]{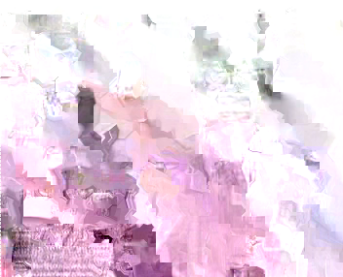}}
		\centerline{(s)}
	\end{minipage}
	\hfill
	\begin{minipage}{0.19\linewidth}
		\centerline{\includegraphics[width=3cm,height=2.5cm]{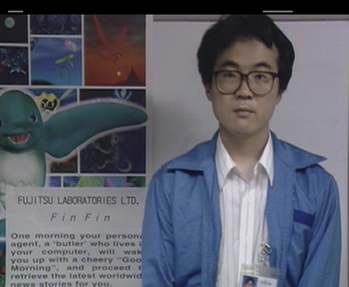}}
		\centerline{(t)}
	\end{minipage}
	\caption{Proposed encryption approach applied to steam Akiyo (\#1 and \#60 frame) and Bowing (\#1 and \#60 frame), demonstrating different amount of visual information in three levels. The first column: the original frames. The second column: encrypted frames with lightweight encryption level. The third column: encrypted frames with Medium encryption level. The fourth column: encrypted frames with heavyweight encryption level. The fifth column: the corresponding decrypted frames of different levels.}
	\label{V}
\end{figure*}
\end{enumerate}

\section{Experimental Results}

In this section, the performance of the proposed multi-level encryption is analysed. A set of benchmark video sequences that are used in the HEVC standardization process are depicted in Figure \ref{IV}. The resolution of the video sequences is 352 x 288, and the frame rate is 60 fps. The sample video frames from the operation of encrypting and decrypting are shown in Figure \ref{IV}. A large number of experiments were performed employing a personal computer configured with an Intel (R) Core (TM) i5 – 4590 CPU @ 2.60 GHz and 16 GB memory, with Windows 10, Visual Studio 2019, MATLAB 2018a, and Opencv 2.4.9. The video coding software HM 16.9 is applied for the proposed scheme. The quantization parameter (QP) is set as 10, 25, and 40.

\begin{figure*}
\centering
\begin{minipage}{0.19\linewidth}
  \centerline{\includegraphics[width=3cm,height=2.5cm]{1.png}}
  \centerline{(a) }
\end{minipage}
\hfill
\begin{minipage}{.19\linewidth}
  \centerline{\includegraphics[width=3cm,height=2.5cm]{2.png}}
  \centerline{(b) }
\end{minipage}
\hfill
\begin{minipage}{.19\linewidth}
  \centerline{\includegraphics[width=3cm,height=2.5cm]{3.png}}
  \centerline{(c)}
\end{minipage}
\hfill
\begin{minipage}{.19\linewidth}
  \centerline{\includegraphics[width=3cm,height=2.5cm]{4.png}}
  \centerline{(d)}
\end{minipage}
\hfill
\begin{minipage}{.19\linewidth}
  \centerline{\includegraphics[width=3cm,height=2.5cm]{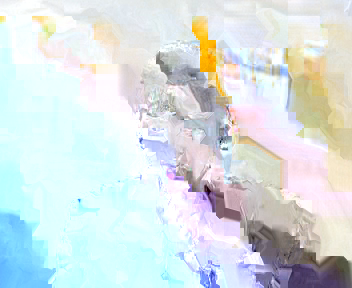}}
  \centerline{(e)}
\end{minipage}

\vfill
\begin{minipage}{0.19\linewidth}
  \centerline{\includegraphics[width=3cm,height=2.5cm]{6.png}}
  \centerline{(f)}
\end{minipage}
\hfill
\begin{minipage}{.19\linewidth}
  \centerline{\includegraphics[width=3cm,height=2.5cm]{7.png}}
  \centerline{(g)}
\end{minipage}
\hfill
\begin{minipage}{.19\linewidth}
  \centerline{\includegraphics[width=3cm,height=2.5cm]{8.png}}
  \centerline{(h)}
\end{minipage}
\hfill
\begin{minipage}{0.19\linewidth}
  \centerline{\includegraphics[width=3cm,height=2.5cm]{9.png}}
  \centerline{(i)}
\end{minipage}
\hfill
\begin{minipage}{0.19\linewidth}
  \centerline{\includegraphics[width=3cm,height=2.5cm]{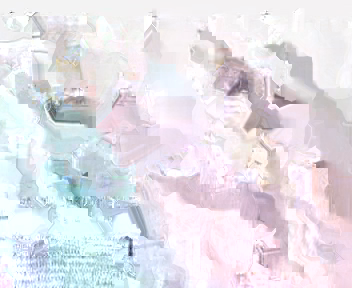}}
  \centerline{(j)}
\end{minipage}
\caption{Comparison of the encryption results of Akiyo (\#1 and \#60 frame) and Bowing (\#1 and \#60 frame) with 4 encryption algorithms. From the first column to fifth column are the original frames, encrypted frames with lightweight encryption level, encrypted frames with medium encryption level, encrypted frames with heavyweight encryption level, encrypted frames with Peng \cite{peng2019tunable}.
}
\label{VV}
\end{figure*}

In Section 4, this paper analyses that the encryption of a certain syntax element can achieve the effect of encryption video with a mass of visual information \cite{Chang2020GuestEI}. According to the amount of visual information, it divides them into three levels to meet the requirements of the users. The major experiments of the proposed scheme include two parts: visual security and encryption security. However, the proposed multi-level encryption is not compared with the state-of-the-art algorithms because the proposed scheme needs to expose some of the visual information and the other algorithms did not reveal any information. The proposed scheme has three encryption levels that have different amounts of visual information. That is, the indexes of different encryption-level experiments should have a sharp division between each other. This paper has performed some related experiments, and the distinction of the experiment’s index has a good fit to the proposed scheme. To illustrate the experiments effect of the three encryption levels, the sequences that are operated by encryption and decryption are shown in Figure \ref{V}. For each video sequence, the encryption effects of I frames and B frames are displayed, respectively, in the first row and the second row of Figure \ref{V}. The performance of Akiyo and Bowing are relatively close in their I and B frames. 

\begin{table*}
\renewcommand\arraystretch{1}
	\footnotesize
	\centering
    \setlength{\tabcolsep}{1.25mm}
	\caption{The average PSNR and SSIM of 60 frames in three levels of algorithms with different QP}\label{runResult}
	\begin{tabular}{lclcccclcccc}
		\hline \hline
		\multirow{2}{*}{\textbf{Video}} & \multicolumn{1}{l}{\multirow{2}{*}{\textbf{QP}}} &  & \textit{\textbf{PSNR}} & \multicolumn{1}{l}{}                                                      & \multicolumn{1}{l}{}                                                 & \multicolumn{1}{l}{}                                                      &                               & \textit{\textbf{SSIM}} & \multicolumn{1}{l}{}                                                      & \multicolumn{1}{l}{}                                                 & \multicolumn{1}{l}{}                                                      \\ \cline{4-7} \cline{9-12}
		& \multicolumn{1}{l}{}                             &  & \textbf{\begin{tabular}[c]{@{}c@{}}Original\end{tabular}}      & \textbf{\begin{tabular}[c]{@{}c@{}}Lightweight\\ Encryption\end{tabular}} & \textbf{\begin{tabular}[c]{@{}c@{}}Medium\\ Encryption\end{tabular}} & \textbf{\begin{tabular}[c]{@{}c@{}}Heavyweight\\ Encryption\end{tabular}} & \multicolumn{1}{c}{\textbf{}} & \textbf{Original}      & \textbf{\begin{tabular}[c]{@{}c@{}}Lightweight\\ Encryption\end{tabular}} & \textbf{\begin{tabular}[c]{@{}c@{}}Medium\\ Encryption\end{tabular}} & \textbf{\begin{tabular}[c]{@{}c@{}}Heavyweight\\ Encryption\end{tabular}} \\ \hline
		\textbf{}                  & 10                                               &  & 50.5860                & 15.7597                                                                   & 11.4146                                                              & 11.2704                                                                   &                               & 0.9946                 & 0.6102                                                                    & 0.4100                                                               & 0.3955                                                                    \\
		\textbf{Akiyo}                       & 25                                               &  & 43.4328                & 15.9601                                                                   & 12.2069                                                              & 11.4938                                                                   &                               & 0.9808                 & 0.6712                                                                    & 0.5112                                                               & 0.5040                                                                    \\
		\textbf{}                       & 40                                               &  & 34.8322                & 14.7331                                                                   & 11.1179                                                              & 10.9469                                                                   &                               & 0.9330                 & 0.6472                                                                    & 0.5237                                                               & 0.5076                                                                    \\\hline
		\textbf{}                 & 10                                               &  & 49.8118                & 13.1044                                                                   & 10.8552                                                              & 10.2631                                                                   &                               & 0.9928                 & 0.5949                                                                    & 0.4589                                                               & 0.4423                                                                    \\
		\textbf{Bowing}                       & 25                                               &  & 44.4147                & 13.0592                                                                   & 11.9812                                                              & 10.7306                                                                   &                               & 0.9859                 & 0.5965                                                                    & 0.5428                                                               & 0.4931                                                                    \\
		\textbf{}                       & 40                                               &  & 34.4788                & 13.0932                                                                   & 10.6565                                                              & 10.3354                                                                   &                               & 0.9352                 & 0.6053                                                                    & 0.5465                                                               & 0.5463                                                                    \\\hline
		\textbf{}               & 10                                               &  & 48.9594                & 13.2268                                                                   & 11.1370                                                              & 11.0232                                                                   &                               & 0.9944                 & 0.4129                                                                    & 0.2471                                                               & 0.2296                                                                    \\
		\textbf{Deadline}                       & 25                                               &  & 40.5181                & 12.8611                                                                   & 11.3648                                                              & 11.1794                                                                   &                               & 0.9781                 & 0.4715                                                                    & 0.3091                                                               & 0.2826                                                                    \\
		\textbf{}                       & 40                                               &  & 30.6408                & 13.5111                                                                   & 12.2058                                                              & 11.9644                                                                   &                               & 0.8832                 & 0.5085                                                                    & 0.3496                                                               & 0.3239                                                                    \\\hline
		\textbf{}                  & 10                                               &  & 48.8314                & 15.8371                                                                   & 10.9286                                                              & 10.9023                                                                   &                               & 0.9906                 & 0.5479                                                                    & 0.3238                                                               & 0.3157                                                                    \\
		\textbf{Irene}                       & 25                                               &  & 41.1426                & 16.2709                                                                   & 12.0858                                                              & 11.9623                                                                   &                               & 0.9705                 & 0.6200                                                                    & 0.4781                                                               & 0.4753                                                                    \\
		\textbf{}                       & 40                                               &  & 32.6193                & 16.2759                                                                   & 11.8398                                                              & 11.7220                                                                   &                               & 0.8828                 & 0.6177                                                                    & 0.4998                                                               & 0.4993                                                                    \\\hline
		\textbf{}                 & 10                                               &  & 49.5047                & 12.4687                                                                   & 12.2353                                                              & 11.3869                                                                   &                               & 0.9922                 & 0.5004                                                                    & 0.4246                                                               & 0.3798                                                                    \\
		\textbf{Mother}                       & 25                                               &  & 42.6266                & 11.8118                                                                   & 10.5153                                                              & 9.5799                                                                    &                               & 0.9730                 & 0.5307                                                                    & 0.5221                                                               & 0.4895                                                                    \\
		\textbf{}                       & 40                                               &  & 34.3846                & 11.8778                                                                   & 11.5669                                                              & 11.1146                                                                   &                               & 0.8822                 & 0.5238                                                                    & 0.5135                                                               & 0.4816                                                                    \\\hline
		\textbf{}                  & 10                                               &  & 48.9832                & 12.2159                                                                   & 11.2651                                                              & 11.2066                                                                   &                               & 0.9951                 & 0.3991                                                                    & 0.2164                                                               & 0.1998                                                                    \\
		\textbf{Paris}                       & 25                                               &  & 39.4438                & 12.5289                                                                   & 11.0768                                                              & 10.9711                                                                   &                               & 0.9776                 & 0.4439                                                                    & 0.2543                                                               & 0.2347                                                                    \\
		\textbf{}                       & 40                                               &  & 29.0746                & 12.7116                                                                   & 11.2934                                                              & 11.2086                                                                   &                               & 0.8689                 & 0.4473                                                                    & 0.2790                                                               & 0.2592                                                                    \\\hline
		\textbf{}                & 10                                               &  & 49.1742                & 10.7477                                                                   & 10.7151                                                              & 10.4624                                                                   &                               & 0.9932                 & 0.4458                                                                    & 0.3646                                                               & 0.3375                                                                    \\
		\textbf{Foreman}                       & 25                                               &  & 40.0126                & 10.9497                                                                   & 10.9346                                                              & 10.8913                                                                   &                               & 0.9612                 & 0.4709                                                                    & 0.4219                                                               & 0.4104                                                                    \\
		\textbf{}                       & 40                                               &  & 31.6714                & 11.2829                                                                   & 10.7100                                                              & 10.4474                                                                   &                               & 0.8646                 & 0.5093                                                                    & 0.4604                                                               & 0.4223                                                                    \\\hline
		\textbf{}               & 10                                               &  & 49.4153                & 13.0439                                                                   & 12.0941                                                              & 11.6432                                                                   &                               & 0.9955                 & 0.5676                                                                    & 0.2491                                                               & 0.2406                                                                    \\
		\textbf{Football}                       & 25                                               &  & 38.6179                & 13.2787                                                                   & 11.9104                                                              & 11.8104                                                                   &                               & 0.9628                 & 0.5882                                                                    & 0.5691                                                               & 0.2554                                                                    \\
		\textbf{}                       & 40                                               &  & 28.1894                & 13.3560                                                                   & 12.2711                                                              & 11.6925                                                                   &                               & 0.7419                 & 0.4888                                                                    & 0.3230                                                               & 0.3061                                                                    \\\hline
		\textbf{}               & 10                                               &  & 49.5828                & 11.9953                                                                   & 11.4926                                                              & 11.1313                                                                   &                               & 0.9946                 & 0.4731                                                                    & 0.2689                                                               & 0.2508                                                                    \\
		\textbf{Pamphlet}                       & 25                                               &  & 43.0321                & 12.5208                                                                   & 10.7428                                                              & 10.6312                                                                   &                               & 0.9851                 & 0.5674                                                                    & 0.4082                                                               & 0.3829                                                                    \\
		\textbf{}                       & 40                                               &  & 33.0452                & 11.8277                                                                   & 12.5812                                                              & 11.3038                                                                   &                               & 0.9053                 & 0.5502                                                                    & 0.4173                                                               & 0.3853                                                                    \\\hline
		\textbf{}              & 10                                               &  & 49.3303                & 11.3327                                                                   & 11.2844                                                              & 10.8286                                                                   &                               & 0.9935                 & 0.5660                                                                    & 0.3623                                                               & 0.3445                                                                    \\
		\textbf{Container}                       & 25                                               &  & 44.2166                & 12.3370                                                                   & 11.3169                                                              & 11.2237                                                                   &                               & 0.9587                 & 0.5825                                                                    & 0.4027                                                               & 0.3914                                                                    \\
		\textbf{}                       & 40                                               &  & 31.5489                & 11.6727                                                                   & 11.2796                                                              & 11.0621                                                                   &                               & 0.8656                 & 0.5660                                                                    & 0.4498                                                               & 0.4398                                                                    \\ \hline
	\end{tabular}
	\label{3}
\end{table*}

\begin{enumerate}[5.1]
\item Analysis of Subjective Vision
\vspace{0.1cm}

To obtain the encryption effect while including some of the visual information, in this paper, the syntax element of the luma IPM, DCT coefficient sign and both of them have been encrypted. The purpose of the proposed method is to provide more selections for video providers and users. There are three levels for them to choose, and each level contains different amounts of visual information that meet the requirements for the video providers and users. Here, Akiyo and Bowing are chosen for analysis.
The proposed scheme has encoded and decoded a video with 60 frames in each encryption level. The decoding video has been depicted into \#1 and \#60 frames, and the encryption effect is shown in Figure \ref{VV}. From Figure \ref{VV}, we distinctly observe that the visual information of the decoded frames after encryption is gradually reduced from the left to right in each rank.
In the lightweight encryption level, one can see the person’s face in the video; moreover, users can obtain a large amount of information. In the medium encryption level, the outline of the object can be easily seen, where one may gain the movement and morphological characteristics of people in the video. In the heavyweight encryption level, it is obvious that we cannot find any visual information from the decoded video after encrypting. Furthermore, it can be found that in each encryption level, the users may obtain different information from the encrypted video. While almost all of the existing HEVC encryption algorithms mainly encrypt the whole video, such that algorithm proposed by Peng \begin{itshape}et al\end{itshape}. \cite{peng2019tunable}, the user without permissions cannot obtain any viewable information. The encryption effect is shown in last column of Figure \ref{VV}.  Therefore, the proposed scheme can meet different requirements for the users.

\begin{table}
\renewcommand\arraystretch{1.1}
	\scriptsize
	\centering
    \setlength{\tabcolsep}{1.5mm}
	\caption{The average entropy of three levels of algorithms}\label{runResult}
	\begin{tabular}{llcccc}
		\hline \hline
		\multirow{2}{*}{\textbf{Video}} &  & \textit{\textbf{ENTROPY}} & \multicolumn{1}{l}{}                                                      & \multicolumn{1}{l}{}                                                 & \multicolumn{1}{l}{}                                                      \\ \cline{3-6}
		&  & \textbf{Original}         & \textbf{\begin{tabular}[c]{@{}c@{}}Lightweight\\ Encryption\end{tabular}} & \textbf{\begin{tabular}[c]{@{}c@{}}Medium\\ Encryption\end{tabular}} & \textbf{\begin{tabular}[c]{@{}c@{}}Heavyweight\\ Encryption\end{tabular}} \\ \hline
		\textbf{Akiyo}                  &  & 7.2205                    & 7.3072                                                                    & 7.4511                                                               & 7.5448                                                                    \\
		\textbf{Bowing}                 &  & 6.7638                    & 6.9155                                                                    & 7.0855                                                               & 7.1504                                                                    \\
		\textbf{Deadline}               &  & 7.2491                    & 7.3753                                                                    & 7.4066                                                               & 7.6384                                                                    \\
		\textbf{Irene}                  &  & 7.0392                    & 7.0725                                                                    & 7.0852                                                               & 7.2087                                                                    \\
		\textbf{Mother}                 &  & 6.8217                    & 7.0302                                                                    & 7.0356                                                               & 7.1084                                                                    \\
		\textbf{Paris}                  &  & 7.1424                    & 7.2265                                                                    & 7.4090                                                               & 7.6469                                                                    \\
		\textbf{Foreman}                &  & 7.4587                    & 7.4848                                                                    & 7.5054                                                               & 7.6170                                                                    \\
		\textbf{Football}               &  & 7.3392                    & 7.3664                                                                    & 7.4268                                                               & 7.4631                                                                    \\
		\textbf{Pamphlet}               &  & 7.1111                    & 7.4066                                                                    & 7.4632                                                               & 7.4852                                                                    \\
		\textbf{Container}              &  & 7.2539                    & 7.3455                                                                    & 7.4429                                                               & 7.5954                                                                    \\ \hline
	\end{tabular}
	\label{4}
\end{table}

\vspace{0.1cm}
\item  Objective Evaluation Index Analysis
\vspace{0.1cm}

To verify the analysis in Section 5.1, the proposed scheme uses the peak signal-to-noise ratio ($PSNR$) and structural similarity ($SSIM$) to measure the performance of the three encryption levels \cite{Wang2004Image}. $PSNR$ is used to measure image distortion, while $SSIM$ measures the similarity of the image. The smaller the values of the two indicators are, the higher the distortion of the frame is, which also means there is less visual information of the frame. Table \ref{3} shows the average $PSNR$ and $SSIM$ of the three encryption levels on 10 videos, in which each video contains 60 frames to ensure that the results are more objective.
From the results, one can see that the value of most sequences is sequentially reduced from the left to the right. It is further proven in the analysis result of subjective vision in Section 5.1, where the visual information is stepped down from the lightweight to the heavyweight encryption level.

\begin{figure}
\renewcommand\arraystretch{1.1}
	\begin{minipage}{0.32\linewidth}
		\centerline{\includegraphics[width=2.5cm,height=2cm]{1.png}}
		\centerline{(a) }
	\end{minipage}
	\hfill
	\begin{minipage}{0.32\linewidth}
		\centerline{\includegraphics[width=2.5cm,height=2cm]{2.png}}
		\centerline{(b) }
	\end{minipage}
	\hfill
	\begin{minipage}{0.32\linewidth}
		\centerline{\includegraphics[width=2.5cm,height=2cm]{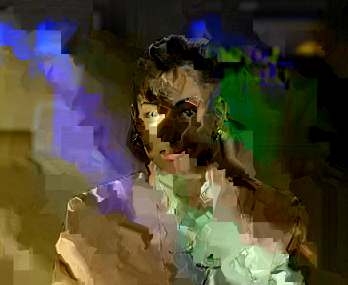}}
		\centerline{(c)}
	\end{minipage}
	
	\vfill
	\begin{minipage}{0.32\linewidth}
		\centerline{\includegraphics[width=2.5cm,height=2cm]{1.png}}
		\centerline{(d) }
	\end{minipage}
	\hfill
	\begin{minipage}{0.32\linewidth}
		\centerline{\includegraphics[width=2.5cm,height=2cm]{3.png}}
		\centerline{(e) }
	\end{minipage}\
	\hfill
	\begin{minipage}{.32\linewidth}
		\centerline{\includegraphics[width=2.5cm,height=2cm]{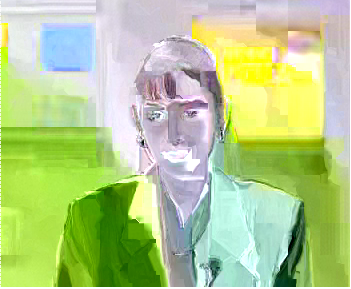}}
		\centerline{(f)}
	\end{minipage}
	
	\vfill
	\begin{minipage}{0.32\linewidth}
		\centerline{\includegraphics[width=2.5cm,height=2cm]{1.png}}
		\centerline{(g) }
	\end{minipage}
	\hfill
	\begin{minipage}{.32\linewidth}
		\centerline{\includegraphics[width=2.5cm,height=2cm]{4.png}}
		\centerline{(h)}
	\end{minipage}
	\hfill
	\begin{minipage}{.32\linewidth}
		\centerline{\includegraphics[width=2.5cm,height=2cm]{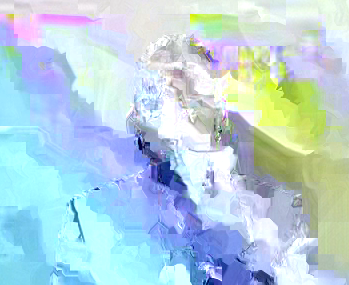}}
		\centerline{(i)}
	\end{minipage}
	\caption{Key sensitivity test of \#1 decoded frame in Akiyo with three encryption levels, demonstrating the security of encryption algorithm. (a) Original frame. (b) lightweight encryption frame. (c) decrypted frame of lightweight encryption with error key. (d) Original frame. (e) medium encryption frame. (f) decrypted frame of medium encryption with error key. (g) Original frame. (h) heavyweight encryption frame. (i) decrypted frame of heavyweight encryption with error key. }
	\label{VI}
\end{figure}

\vspace{0.1cm}
In the video coding process, there are various QP that can be selected, and the smaller the QP is, the more elaborate the coding frame is. As shown in table \ref{3}, with the increasing number of QP in the original video, the $PSNR$ decreases rapidly because the quality of the image is seriously degraded even though one can still see the picture. However, when these video frames have been encrypted, the rank of the $PSNR$ values does not fluctuate too much, and the reason is that when the image is encrypted to a certain extent, the $PSNR$ indicator is not obvious as a measure of the quality of the image, but it can still be used to distinguish the picture with different amounts of visual information. In contrast, the other indicator, $SSIM$, is different than $PSNR$. With the increasing QP number, the value of the $SSIM$ does not change much. However, when the video frame has been encrypted, there is a large change in the $SSIM$. Because $SSIM$ is used to measure the structural similarity between the different video frames, when the video frame has been encrypted, the structure of the video frame is broken, so the value of $SSIM$ extraordinarily changes.

\begin{table*}
	\footnotesize
	\centering
\setlength{\tabcolsep}{1.5mm}
	\caption{The average NPCR and UACI of three levels of algorithms}\label{runResult}
	\begin{tabular}{llcccclcccc}
		\hline \hline
		\multirow{2}{*}{\textbf{Video}} &  & \textit{\textbf{NPCR}} & \multicolumn{1}{l}{}                                                      & \multicolumn{1}{l}{}                                                 & \multicolumn{1}{l}{}                                                      &                               & \textit{\textbf{UACI}} & \multicolumn{1}{l}{}                                                      & \multicolumn{1}{l}{}                                                 & \multicolumn{1}{l}{}                                                      \\ \cline{3-6} \cline{8-11}
		&  & \textbf{Original}      & \textbf{\begin{tabular}[c]{@{}c@{}}Lightweight\\ Encryption\end{tabular}} & \textbf{\begin{tabular}[c]{@{}c@{}}Medium\\ Encryption\end{tabular}} & \textbf{\begin{tabular}[c]{@{}c@{}}Heavyweight\\ Encryption\end{tabular}} & \multicolumn{1}{c}{\textbf{}} & \textbf{Original}      & \textbf{\begin{tabular}[c]{@{}c@{}}Lightweight\\ Encryption\end{tabular}} & \textbf{\begin{tabular}[c]{@{}c@{}}Medium\\ Encryption\end{tabular}} & \textbf{\begin{tabular}[c]{@{}c@{}}Heavyweight\\ Encryption\end{tabular}} \\ \hline
		
		\textbf{Akiyo}                  &  & 0.5370                 & 0.9863                                                                    & 0.9920                                                               & 0.9930                                                                    &                               & 0.0030                 & 0.1795                                                                    & 0.2935                                                               & 0.3000                                                                    \\
		\textbf{Bowing}                 &  & 0.5977                 & 0.9838                                                                    & 0.9914                                                               & 0.9931                                                                    &                               & 0.0034                 & 0.2741                                                                    & 0.2883                                                               & 0.3068                                                                    \\
		\textbf{Deadline}               &  & 0.6571                 & 0.9930                                                                    & 0.9960                                                               & 0.9961                                                                    &                               & 0.0040                 & 0.2139                                                                    & 0.3028                                                               & 0.3143                                                                    \\
		\textbf{Irene}                  &  & 0.8347                 & 0.9927                                                                    & 0.9936                                                               & 0.9953                                                                    &                               & 0.0078                 & 0.2648                                                                    & 0.2772                                                               & 0.3652                                                                    \\
		\textbf{Mother}                 &  & 0.6141                 & 0.9869                                                                    & 0.9886                                                               & 0.9937                                                                    &                               & 0.0042                 & 0.1825                                                                    & 0.3198                                                               & 0.3216                                                                    \\
		\textbf{Paris}                  &  & 0.6959                 & 0.9933                                                                    & 0.9950                                                               & 0.9951                                                                    &                               & 0.0040                 & 0.2480                                                                    & 0.3001                                                               & 0.3030                                                                    \\
		\textbf{Foreman}                &  & 0.6538                 & 0.9960                                                                    & 0.9959                                                               & 0.9961                                                                    &                               & 0.0040                 & 0.3050                                                                    & 0.3002                                                               & 0.3095                                                                    \\
		\textbf{Football}               &  & 0.6382                 & 0.9748                                                                    & 0.9924                                                               & 0.9942                                                                    &                               & 0.0038                 & 0.1687                                                                    & 0.2642                                                               & 0.2817                                                                    \\
		\textbf{Pamphlet}               &  & 0.7939                 & 0.9829                                                                    & 0.9945                                                               & 0.9957                                                                    &                               & 0.0073                 & 0.2202                                                                    & 0.2818                                                               & 0.2964                                                                    \\
		\textbf{Container}              &  & 0.6283                 & 0.9906                                                                    & 0.9953                                                               & 0.9955                                                                    &                               & 0.0036                 & 0.2390                                                                    & 0.2760                                                               & 0.2828                                                                    \\ \hline
	\end{tabular}
	\label{5}
\end{table*}
~\\
\item  Information Entropy Analysis

Information entropy is generally used to measure the information certainty, and it can also apply to the frame \cite{wu2013local,wen2019video}. Larger entropy represents a higher randomness of the distribution of pixels of the whole video frame. To some extent, it can indicate
the security of an encryption algorithm. Therefore, the proposed scheme adopts information entropy to demonstrate the effect of the three encryption levels. The information entropy $ H(I) $ is shown as follows:

 \begin{equation}
H(I)=-\sum\limits _{j=0}^{2^L-1} P(I_j)\log_2 P(I_j).
\end{equation}

where $ L $ represents the number of possible values, and $ P(I_j) $ represents the probability of the pixel value $ I_j $. When all of the possible values have the same probabilities, the information entropy can achieve the maximum value of 8. The closer the entropy of the encrypted image is to 8, the better the encryption performance is. The information entropy of encrypted frames is listed in Table \ref{4}. One can see that the
information entropy value is increasing from left to right, and it indicates that the encrypting performance is gradually increasing. Moreover, from the lightweight encryption level to the heavyweight encryption level, the visual information is gradually reducing. These findings further confirm the analysis results of subjective vision in Section 5.1.

\vspace{0.1cm}
\item Sensitivity Key Analysis
\vspace{0.1cm}

To defend against the brute-force attacks and guarantee the security of the cryptosystem, encryption systems generally should be sensitive to the initial key. Key sensitivity guarantees the uniqueness of the key, and an extremely slight change of the key can lead to completely different results. When the encryption algorithm contains multiple secret keys, a clear plain image cannot be obtained when a key is an error during decryption, and the original clear image can be decrypted only when all the keys are correct. In this experiment, we have picked one key, applied it to the proposed scheme for encryption, and then made a one-bit change in the three encryption levels. The experimental result is demonstrated in Figure \ref{VI}, which shows that there is a great difference in visual information in the decryption process when only one bit has changed in the key. That is, the proposed scheme has a high sensitivity that guarantees the security of the proposed scheme.

\vspace{0.1cm}
\item  $NPCR$ and $UACI$ Analysis
\vspace{0.1cm}

  The functions of the number of pixels change rate ($NPCR$) \cite{hua2016image} and the uniform average change intensity ($ UACI$) \cite{hua2018medical} are used to resist the differential attack. The number of different pixels of two images is measuring by $NPCR$, while $UACI$ collects the different values of pixels of two images. Suppose that $ I_1 $ and $ I_2 $ are two cipher-frames defined as follows:

 \begin{equation}
NPCR(I_1,I_2)=\sum\limits_{n,m}\frac{P(n,m)}{T}\times100\%.
\end{equation}

 \begin{equation}
P(n,m)=\left\{
             \begin{array}{lr}
             0, & if I_1(n,m)=I_2(n,m),\\
             1, & if I_1(n,m)\neq I_2(n,m).\\
             \end{array}
\right.
\end{equation}

\begin{equation}
UACI(I_1,I_2)=\sum\limits_{n,m}{\frac{|I_1(n,m)-I_2(n,m)|}{(L-1)\times T}}\times100\%.
\end{equation}

where $T$ represents the total number of pixels in each cipher-frame, $L$ denotes the number of allowed pixel values, $P$ represents the difference between $ I_1 $ and $ I_2 $, and $ I_1(n,m) $ and $ I_2(n,m) $ indicate the pixel values of two cipher-frames at the position $ (n,m) $. Recently, the expected values of $NPCR$ and $UACI$ are given by

 \begin{equation}\label{special}
NPCR_{expected}=(1-\frac{1}{2^{\log_2 L}})\times100\%.
\end{equation}
 \begin{equation}\label{special2}
UACI_{expected}=\frac{1}{L^2}{\frac{\sum\limits_{v=1}^{L-1} v(v+1)}{L-1}} \times100\%.
\end{equation}
As the test frames are the length of 8-bit pixel value images, the expected $NPCR$ and $UACI$ values are 0.996094 and 0.334635, respectively, according to Eqs. \ref{special} and Eqs. \ref{special2}. When the values of $NPCR$ and $UACI$ are closer to the expected value, the performance of encryption is much better. The $NPCR$ and $UACI$ results of the encrypted frame by the three encryption levels are shown in Table \ref{5}. It can be found that from the lightweight encryption level to heavyweight encryption level, when the $NPCR$ and $UACI$ results of each video are gradually increasing, it means that the ability of resisting the differential attack is becoming increasingly stronger. This finding demonstrates the fact that the performance of the three encryption levels enhances gradually because the visual information is gradually reducing. Another reason is that syntax element of the DCT coefficient sign plays a more important role than the luma IPM, and the encryption performance of encrypting multiple syntax elements is better than that of encrypting a single element.

\begin{figure*}
	\centering
	\begin{minipage}{0.22\linewidth}
		\centerline{\includegraphics[scale=0.27]{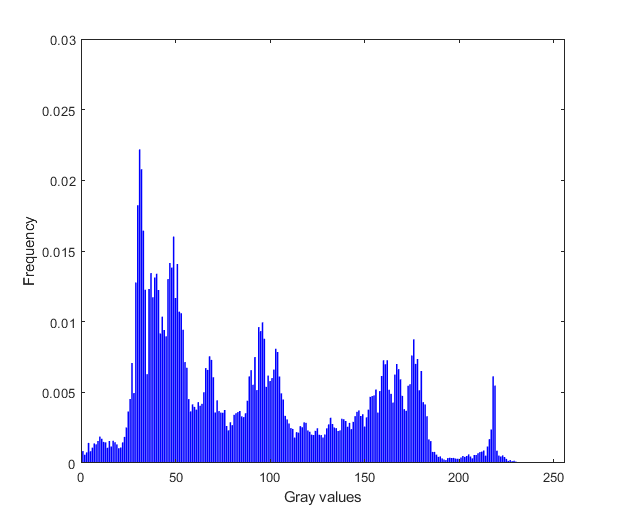}}
		\centerline{(a) }
	\end{minipage}
	\hfill
	\begin{minipage}{0.22\linewidth}
		\centerline{\includegraphics[scale=0.27]{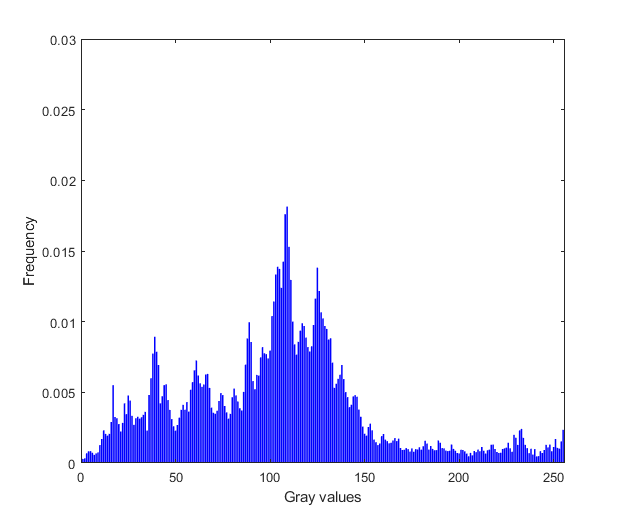}}
		\centerline{(b) }
	\end{minipage}
	\hfill
	\begin{minipage}{0.22\linewidth}
		\centerline{\includegraphics[scale=0.27]{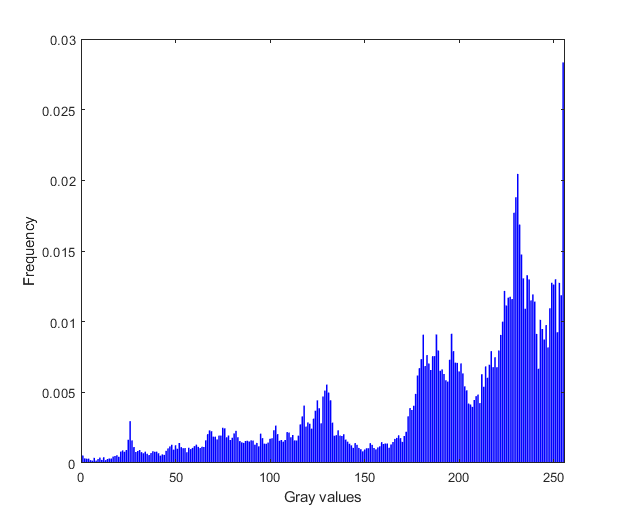}}
		\centerline{(c) }
	\end{minipage}
	\hfill
	\begin{minipage}{0.22\linewidth}
		\centerline{\includegraphics[scale=0.27]{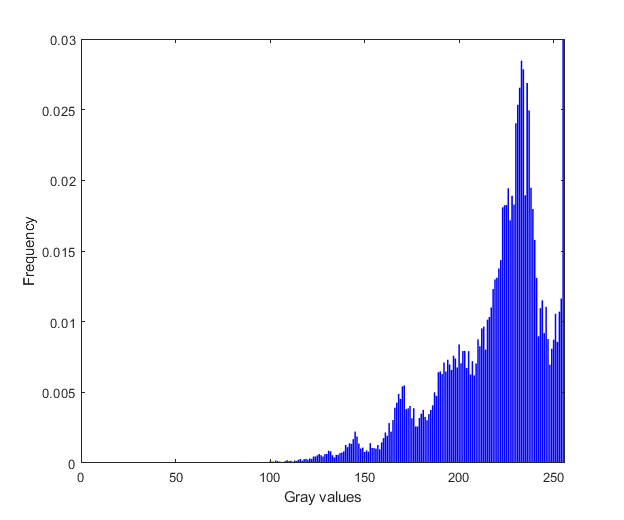}}
		\centerline{(d)}
	\end{minipage}
	\caption{Histogram analysis for Akiyo with three encryption levels, original frames are shown in Fig. 4. (a) histogram of frame in Fig. 4(a). (b) histogram of frame in Fig. 4(b). (c) histogram of frame in Fig. 4(c). (d) histogram of frame in Fig. 4(d).}
	\label{VII}
\end{figure*}

\vspace{0.1cm}
\item  Histogram Analysis
\vspace{0.1cm}

The histogram of a video frame reflects the frequency distribution of pixels. For good performance of encryption, histograms of original and encrypted video frames should differ from each other, and the more different they are, the higher the security of the system is \cite{Sallam2018HEVCSE,ShahidVisual}. However, in the proposed scheme, the experimental results are not entirely different at all. Because in the lightweight encryption level and medium encryption level, the encrypted video frames are not high intensity encryption and there is still a certain residual amount of visual information. The histograms of the video frames encrypted by the three different encryption levels are shown in the Figure \ref{VII}. The histogram of the original video frame, the lightweight encrypted video frame, the medium encrypted video frame, and the heavyweight encryption video frame are presented in Figure \ref{VII} (a), (b), (c), (d), respectively. There are some similarities in the pixel distributions between Figure \ref{VII} (a) and Figure \ref{VII} (b), which means that the video frame that applied the lightweight encryption still reserves a certain amount of visual information. Comparing Figure \ref{VII} (a) and Figure \ref{VII} (c), there is still some resemblance between them; hence, one can obtain some visual information from the video frame after the medium encryption. It is evident from Figure \ref{VII} (a) and Figure \ref{VII} (d) that the histograms of original and encrypted frames are entirely different, which means that the heavyweight encryption has a good encryption performance. It further proves the analysis result of subjective vision in Section 5.1, in which the visual information is stepped down from the lightweight encryption level to the heavyweight encryption level.

\begin{table*}
\renewcommand\arraystretch{1.1}
\setlength{\tabcolsep}{5mm}
\caption{The average bit rate increment of the algorithms  }\label{VVV}
\begin{tabular}{cccccc}
\hline  \hline
\multicolumn{1}{c}{\textbf{\begin{tabular}[c]{@{}c@{}}Original \\ video\end{tabular}}} & \textbf{Akiyo} & \textbf{Bowing}  & \textbf{Deadline} & \textbf{Irene}    & \textbf{Mother}    \\ \hline
\textbf{\begin{tabular}[c]{@{}c@{}}Bit Rate \\ change\end{tabular}}                     & 0.0245         & 0.0200           & 0.0203            & 0.0132            & 0.0136             \\ \hline
\textbf{\begin{tabular}[c]{@{}c@{}}Original\\  video\end{tabular}}                     & \textbf{Paris} & \textbf{Foreman} & \textbf{Football} & \textbf{Pamphlet} & \textbf{Container} \\  \hline
\textbf{\begin{tabular}[c]{@{}c@{}}Bit Rate\\  change\end{tabular}}                     & 0.0202         & 0.0104           & 0.0152            & 0.0193            & 0.0102             \\ \hline
\end{tabular}
\end{table*}

\vspace{0.1cm}
\item  The Security of Key Stream Analysis
\vspace{0.1cm}

In the proposed scheme, AES, as put forward by the U.S. National Institute of Standards and Technology (NIST), is adopted to generate the pseudo-random sequences. This sequence can be considered to have a high level of security because there are no existing algorithms that can break the AES to date. The length of the sequence is more than 192 or 256 bits, which was proven to be secure for protecting the information that needs to be encrypted, even when encrypting a small amount of information.
Furthermore, we can also apply other encryption algorithms such as the Rossler chaotic system \cite{peng2015effective} and 2D logistic-adjusted-sine map \cite{hua2016image} on to generate a pseudo-random sequence for the proposed scheme. That is, there is nothing specific to the encrypted content to consider; the security of the scheme depends on the security of the algorithm.
In this paper, we pay more attention to the encryption performance of different syntax elements in encrypted H.265/HEVC, so the security of the encryption algorithm that we adopted is not discussed and tested in detail in this paper.

\vspace{0.1cm}
\item  Bit Rate Change Analysis
\vspace{0.1cm}

In video encryption algorithm, one of a significant index is bit rate change \cite{Dong2018AccurateDO}. Keeping the video bit rate is an ideal state for video encryption. In general, the encrypted syntax elements in bypass mode can keep the big rate while the bit rate is inevitably increased as long as the syntax elements encoded in regular mode. In the proposed scheme, the encrypted syntax elements of the DCT coefficient sign and luma IPM are in the bypass mode and regular mode, respectively. There is one encrypted syntax element of luma IPM in the regular mode, therefore we only need to calculate the bit rate change in lightweight encryption level or heavyweight encryption level. The average bit rate increment is listed in Table \ref{VVV}. It can be found that the bit rate change is slightly increment, almost under 2.5\%. The result demonstrates that the encryption algorithm is almost not impacted video compression coding system.

\end{enumerate}

\section{Conclusions}

In this paper, we learn about the real-world requirements of video encryption and analyse some contradictions between users and video providers. To meet the various requirements
of users and video providers, as a benefit for both of them, a multi-level selective encryption scheme for H.265/HEVC is proposed based on encrypting syntax elements in the coding process. There are three levels for users, which are the lightweight encryption level, the medium encryption level, and the heavyweight encryption level. The syntax element of luma IPM is encrypted in the lightweight encryption level, the syntax element of the DCT coefficient sign is chosen for encryption in the medium encryption level, and both of the syntax elements are encrypted in the heavyweight encryption level. Since only a few numbers of syntax elements are encrypted, the users can always gain some visual information from the encryption videos. The experimental results and analysis confirm that the amount of visual information contained in the three encryption levels is different, and the visual information is reduced successively from the lightweight encryption to the heavyweight encryption, which exactly positions the approach as a solution for the contradiction between supply and demand that exists between users and video providers.

\bibliographystyle{elsarticle-num-names}
\bibliography{bibfile}

\end{document}